\documentclass[%
 aip,
 amsmath,amssymb,
 reprint,%
]{revtex4-1}

\usepackage{graphicx}
\usepackage{dcolumn}
\usepackage{bm}
\usepackage{textcomp}

\usepackage[utf8]{inputenc}
\usepackage[T1]{fontenc}
\usepackage{mathptmx}
\usepackage{soul} 

\soulregister\ref{7}
\soulregister\eqref{7}
\soulregister\cite{7}
\soulregister\onlinecite{7}

\begin{document}

\preprint{AIP/123-QED}

\title[Creating and Controlling Complex Light]{Creating and Controlling Complex Light}

\author{Nicholas Bender}
\altaffiliation{Department of Applied Physics, Yale University, New Haven, CT 06511, USA}

\author{Hasan Y{\i}lmaz}
\affiliation{Department of Applied Physics, Yale University, New Haven, CT 06511, USA}

\author{Yaron Bromberg}
\affiliation{Racah Institute of Physics, The Hebrew University of Jerusalem, Jerusalem 91904, Israel}

\author{Hui Cao}
\affiliation{Department of Applied Physics, Yale University, New Haven, CT 06511, USA}
\email{hui.cao@yale.edu}

\date{\today}

\begin{abstract}
Random light fields -commonly known as speckles- demonstrate Rayleigh intensity statistics and only possess local correlations: which occur within the individual speckle grains. In this work, we develop an experimental method for customizing the intensity probability density function (PDF) of speckle patterns while simultaneously introducing non-local spatial correlations among the speckle grains. The various families of tailored speckle patterns -created by our method- can exhibit radically different topologies, statistics, and variable degrees of spatial order. Irrespective of their distinct statistical properties, however, all of these speckles are created by appropriately encoding high-order correlations into the phase front of a monochromatic laser beam with a spatial light modulator. In addition to our experimental demonstration, we explore both the theoretical and practical limitations on the extent to which the intensity PDF and the spatial intensity correlations can be manipulated concurrently in a speckle pattern. This work provides a versatile methodology for creating complex light fields and controlling their statistical properties with varied applications in microscopy, imaging, and optical manipulation.
\end{abstract}

\maketitle

\section*{Introduction}
Spatially random light fields have the hallmark appearance of intricate -yet highly irregular- mosaics of diffraction-limited speckle grains. Because of their speckled appearance random light fields are commonly referred to as speckle patterns. A speckle pattern is characterized by the twofold complexity of its optical field. On one hand, the spatial-distribution of light in a speckle pattern is sufficiently complicated that speckles are described by a statistically stationary and ergodic random process. In this context, stationarity requires the statistical properties of an ensemble of speckle patterns to be the same as those of an individual speckle pattern within the ensemble. Ergodicity requires the statistical properties of two spatial positions -separated by more than one speckle grain size- to be independent and identical to those of the ensemble.  On the other hand, speckle patterns are categorized by the joint PDF of their complex-valued field. For example, a speckle pattern is said to be ‘fully developed’ if its joint PDF is circularly invariant. In a fully-developed speckle pattern, therefore, the phase PDF is uniformly distributed between $0$ and $2 \pi$. Additionally, in a fully-developed speckle pattern the amplitude and phase profiles are statistically independent. Rayleigh speckles -the most common type of speckle patterns- obey a circular-Gaussian field PDF which results in a negative exponential intensity PDF. Furthermore, they only possess short-ranged spatial intensity correlations which are determined by the average speckle grain shape: which in turn is dictated by the diffraction limit. \cite{GoodmanB, Dainty1, DaintyB, Freund1001}

Over the years, various methods have been developed to modify the intensity statistics of speckle patterns. For example, it has been shown that the intensity PDF of a speckle pattern can be made non-Rayleigh, \cite{Dainty2, Dainty3, Goodman2, Asakura1, Asakura2, Jakeman1, Jakeman2, Pedersen, ODonnell} however, the resulting speckle pattern is typically either under-developed or partially-coherent, which only allows for a limited range of possible functional forms for the speckles’ intensity PDF. Nevertheless, recent works have shown that it is possible to modify the intensity PDF of a fully-developed speckle pattern.\cite{Yaron, CSS, Guillon17,JesusSilva, ZNie, PHong} In these works, other statistical properties such as the functional form of the speckles’ spatial intensity correlations remain unchanged.

In a speckle pattern, the spatial field correlation function is defined as:
\begin{equation} 
C_{E}(\Delta {\bf r}) \equiv {\langle E({\bf r}) E^{*}({\bf r} + \Delta {\bf r})\rangle}/{\langle |E({\bf r})|^2\rangle}
\end{equation}
where $\langle ... \rangle$ denotes spatial averaging over $\bf r$. The spatial intensity correlation function is given by:
\begin{equation} 
\begin{split}
C_{I}(\Delta {\bf r}) & \equiv {\langle I({\bf r}) I({\bf r} + \Delta {\bf r})\rangle}/{\langle I({\bf r}) \rangle \langle I({\bf r} +\Delta {\bf r}) \rangle} -1\\
& = C_{L} (\Delta {\bf r}) + C_{NL}(\Delta {\bf r}).
\end{split}
\end{equation}
Here $C_{L} (\Delta {\bf r})$ is known as the local (short-range) correlation function, and it is related to the field correlation function by $C_{L}(\Delta {\bf r}) = C_{0} |C_E(\Delta {\bf r})|^2$, where $C_{0} = \langle I^{2} \rangle /\langle I \rangle^{2} -1$ is related to the speckle contrast. \cite{Freund1001, GoodmanB, DaintyB, Mello2} $C_{NL}(\Delta {\bf r})$ represents the non-local (long-range) correlation function,\cite{BerkovitsPR94} and it vanishes when the Siegert relation holds: $C_{I}(\Delta {\bf r}) \equiv C_{0} |C_E(\Delta {\bf r})|^2$. Typically, the spatial intensity correlation function of a speckle pattern can be modified by altering the local correlation function: via amplitude modulation of its Fourier components.~\cite{Asakura1, FractalSpec, ICorManip, Markovian, chriki2018rapid, waller2012phase, 2012_Fleischer_PRL, Battista, Phillips, IntroNonLocal, Sheridan, Eimerl, YoshimuraJOSAA92}  Because the local correlation function is effectively the diffraction-limited point spread function of a system, this approach can be quite limiting in terms of the range of possible correlation functions. In our recent work,\cite{IntroNonLocal} we demonstrated that it is possible to dramatically and controllably alter the intensity correlation function of a speckle pattern by introducing non-local correlations into the speckle pattern instead. While this method could modify the speckle contrast, via $C_{0}$, the intensity PDF itself could not be directly controlled using this method.

In this article, we experimentally demonstrate a method of \textit{simultaneously} customizing the intensity PDF of speckle patterns and introducing long-range spatial correlations among the speckle grains. Various families of speckles are created by encoding high-order correlations into the phase front of a monochromatic laser beam with a spatial light modulator (SLM): using a two-stage method.  In addition to our experimental demonstration, we explore both the theoretical and practical limitations on the extent to which the intensity PDF and the spatial intensity correlations can be manipulated simultaneously in a speckle pattern without modifying the spatial field correlation function.

The ability to independently control the intensity PDF and correlations of speckles -arbitrarily- has many potential applications. For example they can be used as a form of `smart' illumination in high-order ghost imaging, \cite{image1,image2, image3}  dynamic speckle illumination microscopy, \cite{Dynamic2, Dynamic4} super-resolution imaging, \cite{2012_Sentenac, Super_opt1,  2014_Zheng_super-resolution, 2017_Waller_super-resolution, Super_acous1} compressive sensing, \cite{DogariuOptica17, 2019_Guillon_NatCommun} and optical sectioning microscopy.  \cite{mertz2011optical} Furthermore, using speckle patterns with customized intensity statistics as bespoke disordered optical potentials in transport studies of cold atoms, \cite{cold} colloidal particles, \cite{coll} and active media \cite{active} could induce novel transport behaviors.

\section*{Materials and Methods}

\subsection*{Experimental Setup}
\begin{figure}[hthb]
    \centering
    \includegraphics[width=\linewidth]{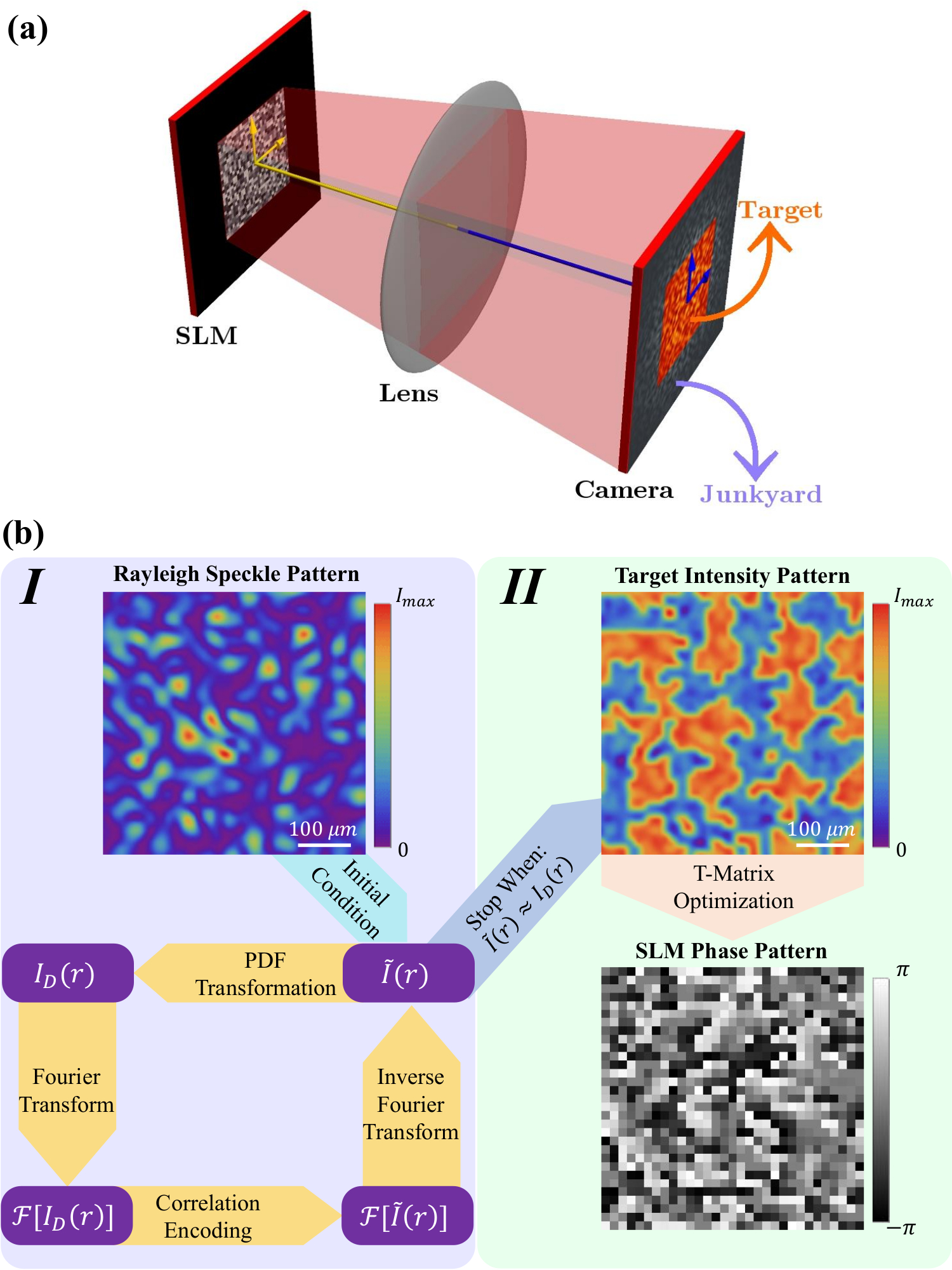}
    \caption{In (a) we present a schematic illustration of the speckle creation setup where part of a phase-only SLM is illuminated by a laser beam and the speckle pattern in its Fourier plane is recorded by a camera. In (b) we show a flowchart visualization of our speckle customization method. $I_D(\bf r)$ represents an intensity pattern obeying the desired PDF and $\tilde{I}(\bf r)$ is an intensity pattern with the desired correlations.}
    \label{Figure0}
\end{figure}

In our experimental setup, illustrated in Figure~\ref{Figure0} (a), a linearly-polarized monochromatic laser beam with a wavelength of $\lambda = 642$ nm uniformly illuminates a phase-only reflective SLM (Hamamatsu LCoS X10468). The pixels on the SLM can modulate the incident light’s phase between the values of $0$ and $2 \pi$ in increments of $2 \pi / 170$. However, a small portion of reflected light from the SLM is unmodulated. To bypass the unmodulated light, we write a binary phase diffraction-grating on the SLM, and work with the light diffracted to the first-order. In order to avoid crosstalk between neighboring SLM pixels, $16 \times 16$ pixels are grouped to form one macropixel, and the binary diffraction-grating is written within each micropixel: with a period of 8 pixels. In order to use the Hadamard basis in the measurement of the field transmission matrix, we use a square array of $32 \times 32$ macropixels in the central part of the phase modulating region of the SLM. Outside the central square, the remaining illuminated pixels display an orthogonal phase grating to diffract the laser beam away from the CCD camera (Allied Vision Prosilica GC660). The SLM and CCD camera are placed on opposing focal planes of a lens ($f=500$ mm). To a good approximation the field incident on the camera is a Fourier transform of the field reflected off the SLM. To be more precise and general, however, we use an experimentally measured field-transmission matrix (T-matrix) to relate the light field on the SLM and the camera planes. This enables us to easily account for experimental artifacts: such as curvature on the SLM surface, lens aberrations, and misalignments in the optical system.

In our setup, the T-matrix is measured using a self-interference method.~\cite{Poppoff,Yoon} The notable difference between our measurement technique, and those cited,~\cite{Poppoff,Yoon} is that we superimpose a random phase pattern on the SLM when performing a T-matrix measurement. This enables us to average over multiple T-matrix measurements -using different superimposed random phase patterns on the SLM- and significantly reduce the error of our final T-matrix. Therefore, for a given phase pattern displayed on the SLM, the average differences between the speckle intensity pattern measured by the CCD camera and that predicted by the field transmission matrix is less than $10\%$. 

\subsection*{Method}

Our method of creating and controlling complex light, by simultaneously controlling the intensity PDF, $P(I)$, and the spatial intensity correlation function, $C_I(\Delta {\bf r})$, in a speckle pattern has two fundamental stages as depicted in Figure~\ref{Figure0} (b). First, a target speckle intensity pattern which obeys the desired intensity statistics, both $P(I)$ and $C_I(\Delta {\bf r})$, is numerically generated by transforming a Rayleigh speckle pattern. Once a target intensity pattern is known, the next stage involves using a field-transmission-matrix based nonlinear optimization algorithm to obtain a corresponding speckle field -possessing the desired target intensity profile- which can be created in our experimental setup using a phase-only SLM. Repeating our method with different/uncorrelated initial Rayleigh speckle patterns enables us to create a stationary and ergodic ensemble of speckle patterns obeying the desired custom statistics.

Irrespective of the optimization method used to generate a phase pattern on the SLM -which creates customized speckles on the camera plane- the problem is non-convex and the search’s parameter-space is vast. Therefore, while it is possible to directly search for a SLM phase pattern which generates a customized speckle pattern with the desired statistical properties -without using a target intensity pattern- this approach is not necessarily ideal. For example, if such a method fails to converge to an acceptable solution it would be difficult determine if this was because the algorithm was not optimal or if the desired statistics were fundamentally impossible to encode into a speckle pattern. By partitioning our method into two steps -first generating a speckle intensity pattern with the desired statistics and then creating the speckle pattern with the SLM- this can be differentiated. Furthermore, this division reduces the parameter space of our search for a solution and enables us to use a \textit{local}, as opposed to a \textit{global}, search algorithm. As a result our algorithm always converges to an acceptable solution in a reasonable amount of time.

\subsubsection*{Designing Custom Speckles}

To complete the first stage of our method, we generate a speckle intensity pattern, $I({\bf r})$, which adheres to a desired intensity probability density function, $P(I)$, and has a spatial intensity correlation function, $C_I(\Delta {\bf r})$, with a tailored functional form: by transforming an experimentally measured Rayleigh speckle pattern. In order to successfully encode both desired statistical properties into our target intensity pattern, we use an individual transformation for each property: using our previously developed methods for customizing either $P(I)$ or $C_I(\Delta {\bf r})$. \cite{CSS, IntroNonLocal} To begin with, we can modify the intensity PDF of a speckle intensity pattern by performing a local intensity transformation on it, as shown previously.\cite{yura2012digital, CSS} In general, a local intensity transformation is defined such that if  $I_{0}({\bf r})$ is an initial speckle intensity pattern adhering to the intensity PDF, $P_{0}(I)$, then the scalar transformation $f(I_{0}({\bf r})) = I_{D}({\bf r})$ will produce a new intensity pattern $I_{D}({\bf r})$ which adheres to the desired intensity PDF, $P_{D}(I)$. The specific local intensity transformation associated with the target PDF, $P_{D}(I)$, can be found from the integral
\begin{equation}
\label{int}
\int_{0}^{I_{0}} P_{0}(I') dI' =\int_{0}^{I_{D}} P_{D}(I') dI' .
\end{equation}
By expressing $I_{D}$ as a function of $I_{0}$, we obtain the local intensity transformation $f(I_{0}) = I_{D}$. While this enables us to customize the intensity PDF of a speckle pattern, long-range spatial intensity correlations are not modified by this operation. \cite{IntroNonLocal} To introduce the desired $C_I(\Delta {\bf r})$ into a speckle intensity pattern,  we employ the relation between the  Fourier transform of a speckle intensity pattern $\mathcal{F}[I({\bf r})]$ and its spatial intensity correlation function $C_{I}(\Delta {\bf r})$: $\mathcal{F}[C_{I}(\Delta {\bf r})+1]=| \mathcal{F}[I({\bf r})]|^{2}$. Therefore by modifying a speckle intensity pattern’s Fourier amplitude according to $|\mathcal{F}[I({\bf r})]| = |\sqrt{\mathcal{F}[C_{I}(\Delta {\bf r})+1]}|$, we can create a speckle intensity pattern, $\tilde{I}({\bf r})$, which obeys the desired intensity correlation function. \cite{IntroNonLocal} Because information related to customizing the intensity PDF is encoded into the spatial representation of a speckle pattern -via the local intensity transformation of $I({\bf r})$- and the desired intensity correlation function information is encoded in the Fourier representation of the speckle intensity pattern -by imposing $|\sqrt{\mathcal{F}[C_{I}(\Delta {\bf r})+1]}|$- we can merge both customization methods into a single Gerchberg-Saxton algorithm as illustrated in Figure~\ref{Figure0} (b). In this process, the only constraint on the statistical properties encoded into the speckle pattern is the fundamental relationship between the speckle correlation function and the intensity PDF: $C_{I}(0) = C_{0} = \langle I^{2} \rangle /\langle I \rangle^{2} -1$. Beyond this, however, the functional form of $C_{I}(\Delta {\bf r})$ may be chosen independently from $P(I)$.
 
The first step of our Gerchberg-Saxton algorithm is to perform a local intensity transformation on a Rayleigh speckle pattern which converts it into a speckle intensity pattern, $I_{D}({\bf r})$, governed by the desired PDF. Next, we modify the amplitude of its Fourier components, such that $|\mathcal{F}[I_{D}({\bf r})]|$ is equal to the desired $|\sqrt{\mathcal{F}[C_{I}( \Delta{\bf r})+1]}|$, without altering the phase values. The inverse Fourier transform of the modified Fourier spectrum gives a complex valued function for the intensity pattern, $\tilde{I}({\bf r})$, which obeys the desired correlation function. Since the intensity values must be positive real numbers, we ignore the phase values and set $\tilde{I}({\bf r}) = |\tilde{I}({\bf r})|$. In the process of encoding correlations into $I_{D}({\bf r})$, the intensity PDF that the resulting pattern, $\tilde{I}({\bf r})$, obeys is altered slightly relative to that of $I_{D}({\bf r})$. This deviation from the desired PDF is corrected for by applying an appropriate local intensity transformation to $\tilde{I}({\bf r})$. While the modified intensity pattern now obeys the desired speckle intensity PDF, the application of a local intensity transformation to $\tilde{I}({\bf r})$ has slightly altered the spatial intensity correlations previously encoded into the speckle pattern. The small deviation from the desired $C_{I}( \Delta{\bf r})$ can be corrected by resetting the Fourier amplitude of the intensity pattern to $|\sqrt{\mathcal{F}[C_{I}( \Delta{\bf r})+1]}|$. Cyclical repetition of this process results in an intensity pattern which adheres to both the desired correlation function and the PDF: therefore we have $I_{D}({\bf r})=\tilde{I}({\bf r})$.  Starting with different initial Rayleigh speckle patterns produces uncorrelated intensity patterns that satisfy the same $C_I(\Delta {\bf r})$ and $P_{D}(I)$, and therefore by using a stationary and ergodic ensemble of uncorrelated Rayleigh speckle patterns we can create a stationary and ergodic ensemble of uncorrelated customized speckle patterns.

Although this method excels at generating speckle patterns when the desired non-local correlation function has sparse Fourier components, it may converge to an ordered -as opposed to speckled- intensity pattern when the desired non-local correlation function is sparse in real space -therefore dense in Fourier space- such as the example shown in Figure~\ref{Figure1}. In this case, rather than producing a random intensity pattern, the Gerchberg-Saxton algorithm converges to an ordered pattern which adheres to the desired intensity PDF and $C_I(\Delta {\bf r})$. As was shown in reference,~\cite{IntroNonLocal} we can rectify this absence of disorder by convolving the ordered intensity pattern with a speckle pattern which does not possess any long-range intensity correlations, such as a Rayleigh speckle pattern. This convolution does not alter the functional form of $C_I(\Delta {\bf r})$ since Rayleigh speckles only have short-ranged correlations, however, it may alter the speckles' intensity PDF. In general, this alteration can be removed by using the convolved speckle pattern as the initial speckle pattern of a second Gerchberg-Saxton algorithm: which follows the same procedure as the first Gerchberg-Saxton algorithm. In the event that $P(I)$ is smooth the alteration of $P(I)$ is minor~\cite{CSS} and to a good approximation only the value of $C_I(0)$ changes: \textit{i.e.} the variance of the encoded PDF. In this case, we can use either a super-Rayleigh or sub-Rayleigh speckle pattern -which have varying intensity contrasts and only local correlations \cite{Yaron}- in the convolution to adjust the value of $C_I(0)$: as was done for the case shown in Figure~\ref{Figure1}. 

\subsubsection*{Creating Custom Speckles}

The second stage of our method consists of using a T-matrix based nonlinear-optimization algorithm to determine the phase pattern which -upon application to the SLM- generates a desired speckle intensity pattern on the CCD camera plane. The reason we resort to an optimization algorithm, as opposed to an analytical expression, is that the phase values of the $(32\times 32)$ macropixels on the SLM are transcendentally related to the intensity values measured by the CCD camera. To find a solution for the SLM phase array which generates a given target intensity pattern, we numerically minimize the difference between the target pattern, $\tilde{I}({\bf r})$, and the intensity pattern, $I_{M}({\bf r})$, obtained after applying the field transmission matrix to the SLM phase array: as a function of the SLM phases. Specifically, $ I_{M}({\bf r})= |\sum_{n} t_{n} ({\bf r}) {\rm e}^{{\rm i} \theta_{n}}|^2$, where $\theta_{n}$ represents the phase displayed on the $n^{\rm th}$ SLM pixel and $t_{n} (\bf{r})$ the element of the transmission matrix mapping the $n^{\rm th}$ SLM pixel to the $\bf{r}$  position on the camera. The cost function $\sum_{{\bf r}} | \tilde{I}({\bf r})- I_{M}({\bf r})|^{2}$ is minimized by tuning the SLM phase $\theta_n$. To facilitate the convergence to a solution, we reduce the area we attempt to control -on the camera plane- to the central quarter region representing the Fourier transform of the phase modulating region of the SLM.
As shown previously, \cite{CSS, IntroNonLocal} in order to avert the effects of aliasing and uniquely define the spatial profile of a speckle intensity pattern, it is necessary to sample the pattern at or above the Nyquist limit. Therefore, each speckle grain should be sampled at least twice along both spatial axes. This means that the $(32\times 32)$ speckle grains measured by our CCD camera must be represented by $(64\times 64)$ partially correlated intensity values. By reducing our target region to only containing $(16\times 16)$ speckle grains -therefore only $(32\times 32)$ partially correlated intensity values need to be controlled by $(32\times 32)$ independent phase values- we guarantee the existence of multiple SLM phase arrays for a given target intensity pattern. Because the ($32 \times 32$) intensity values are partially correlated, the number of degrees of freedom in the target plane is effectively less than the number of degrees of freedom we have available on the SLM. Due to this we can use a local-search algorithm, the Broyden-Fletcher-Goldfarb-Shanno (BFGS) algorithm, \cite{nLopt2} to solve for the SLM phase array and will always obtain a solution in a relatively short amount of time. For example, using \textit{Matlab} on a laptop with an \textit{ Intel} i7-4910MQ processor (2.9 GHz base frequency), it takes about 45s to obtain a SLM phase array which generates a speckle pattern with the desired target intensity PDF and spatial correlation function. In the process of solving for a SLM phase array, our local-search algorithm appropriately encodes high-order correlations into the phase values of the SLM phase array. \cite{CSS,IntroNonLocal}

\begin{figure}[hthb]
    \centering
    \includegraphics[width= \linewidth]{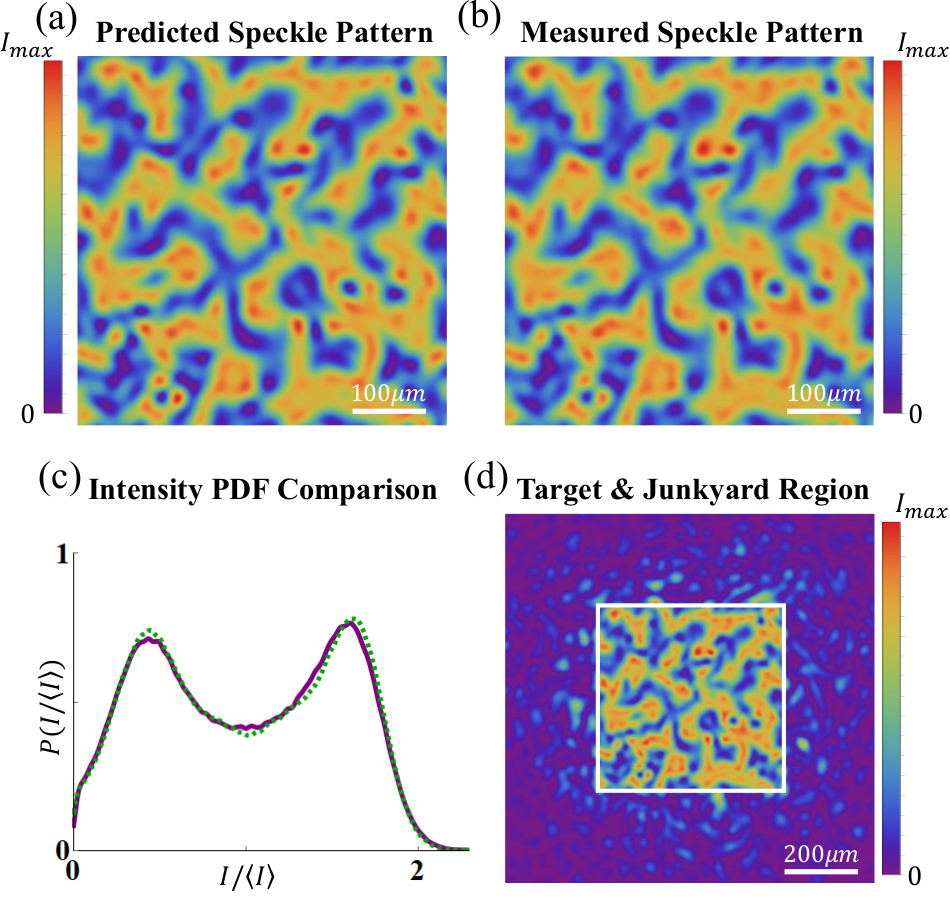}
    \caption{An example customized speckle pattern predicted by our T-matrix (a) is juxtaposed with the corresponding experimentally measured speckle pattern (b). The difference between the two intensity patterns is $9.7\%$. In (c) we compare the intensity PDF of the predicted speckle patterns, green dashed line, with the intensity PDF of the corresponding measured speckle patterns, purple solid line, for an ensemble of 100 speckles patterns like those shown in (a) and (b). The difference between the two intensity PDFs is $3.7\%$. In (d) we present a measured image of the speckles in both the target region and the junkyard region. The white square denotes the boundary of the target region.}
    \label{FigureTM}
\end{figure}

To check the error of our method/system, we compare a customized speckle pattern predicted by the measured T-matrix, in Fig.~\ref{FigureTM}(a), with the corresponding experimentally measured speckle pattern, in Fig.~\ref{FigureTM}(b). The difference between the two intensity patterns is $9.7\%$, which is typical. Because the customized properties of the speckle patterns are statistical - both the intensity PDF and the spatial intensity correlation function- they are robust to minor differences between the measured and predicted speckle patterns. For example, in (c) we compare the intensity PDF of the predicted speckle patterns, green dashed line, with the intensity PDF of the corresponding measured speckle patterns, purple solid line, for an ensemble of 100 speckles patterns like those shown in (a) and (b). The difference between the two intensity PDFs is just $3.7\%$: less than the difference between the two speckle patterns. This is because the averaging inherent to calculating the respective intensity PDFs suppresses, rather than compounds, the effects of fluctuations/deviations between the two patterns.

The speckle pattern shown in Fig.~\ref{FigureTM}(b) is located within the target region. However, the speckles in the region outside of it -which we call the junkyard- have distinct statistical properties relative to those in the target region. Fig.~\ref{FigureTM}(d) is a measured speckle pattern including both the target region and the junkyard. The image encompasses the complete Fourier plane of the SLM. While the central target region (denoted by the white square) adheres to the desired intensity PDF and spatial intensity correlations, the speckles in the junkyard region do not. Though the precise statistical properties of the speckles in the junkyard region depend on the details of the target region's speckles, they approximately adhere to Rayleigh statistics and are devoid of non-local correlations.

\section*{Customized speckles and high-order statistics}

\begin{figure}[hthb]
    \centering
    \includegraphics[width= \linewidth]{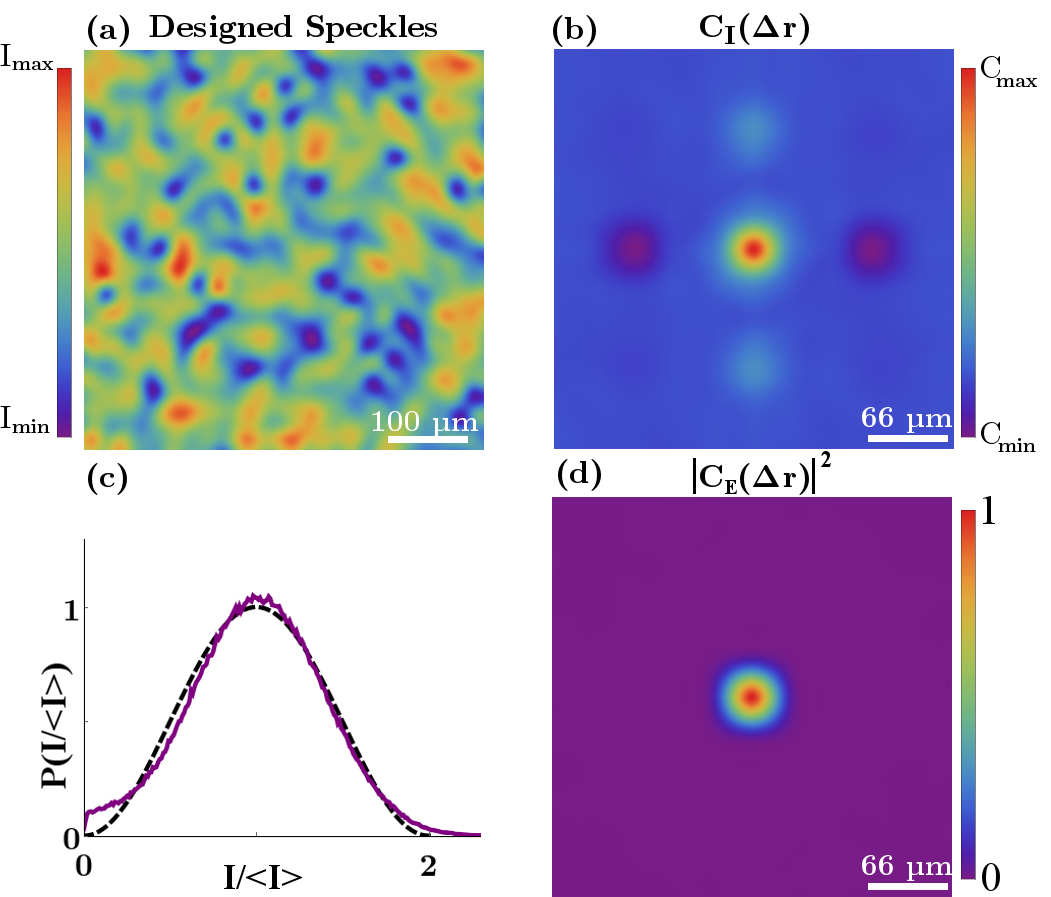}
    \caption{ A customized speckle pattern is shown in (a) with a spatially sparse intensity correlation function (b) and a unimodal intensity PDF (c). In (b) $C_{\rm max}=0.15$ and $C_{\rm min}=-0.03$. In (c) the intensity PDF of the experimentally created speckles (purple solid line) closely follows the target intensity PDF (black dashed line). The local intensity correlation function (d) remains the same as that of a Rayleigh speckle pattern, indicating that the modification of the spatial intensity correlation function is the result of introducing non-local correlations into the speckle pattern. To obtain (b-d) we ensemble average over 100 independent speckle patterns to obtain the PDF and the correlation functions. The origin of (b) and (d) is located at the center of each plot.}
    \label{Figure1}
\end{figure}

In Figure~\ref{Figure1} (a) we present an example of an experimentally measured speckle pattern which is customized to simultaneously possess the spatial intensity correlation function shown in (b) and adhere to the intensity PDF shown in (c). The experimentally obtained intensity PDF (purple line) in (c) was tailored to have the form, $P(\hat{I})=\sin^{2} [{\pi \hat{I}}/{2}]$, over the range, $0 \leq \hat{I} \equiv I / \langle I \rangle \leq 2$,  and $P(\hat{I})=0$ for values of $\hat{I}>2$ (black dashed line). The two curves closely follow one another, except around $\hat{I}=0$. This deviation occurs because optical vortices are inherently present in the experimentally-generated speckle patterns, and therefore the measured probability around $\hat{I}=0$ must be nonzero. The intensity correlation function shown in (b) was designed to have positive correlations, $C_{I}(\Delta {\bf r}) = 0.03$, at $\Delta {\bf r} = (0, \pm 100$ \textmu m$)$  and negative correlations, $C_{I}(\Delta {\bf r}) = -0.03$, at $\Delta {\bf r} = ( \pm 100$ \textmu m$)$. Because our method is based on the use of a phase-only SLM, which is in the Fourier plane of our camera, the Fourier amplitude profile of the speckle fields generated in the CCD camera plane is fixed. Therefore, due to the Wiener-Khinchin theorem, the spatial field correlation function of a customized speckle pattern in the camera plane remains identical to that of the initial Rayleigh speckle pattern (created by a random phase array on the SLM): as demonstrated in (d). Because our modification of the speckle pattern’s intensity correlation function does not affect the field correlation function, and therefore the local intensity correlations, the modified intensity correlations are non-local.~\cite{IntroNonLocal}

Often, complex-light patterns are classified in terms of a defining statistical property, such as the intensity PDF: the most common example would be a \textit{Rayleigh} speckle pattern. Such a characterization requires the existence of an ensemble/family of independent speckle patterns which individually adhere to the stated statistical property. The speckle patterns generated using our method are no exception; while we present the intensity PDF and spatial correlation function calculated using the entire ensemble of speckle patterns -in Figure~\ref{Figure1}- each speckle pattern adheres the stated statistical properties individually and therefore is part of statistically stationary and ergodic ensemble. We can verify that the intensity PDFs of the speckle patterns are stationary by calculating the average deviation of the PDF of a single speckle intensity pattern, $P_{S}(I)$, from the intensity PDF constructed using the  ensemble of speckle patterns, $P_{E}(I)$. We quantify the difference between the PDFs using the formula: $\Delta P_{S} = \left[ \langle |P_{E}(I) - P_{S}(I)| \rangle_{I} \right] / \left[ \sqrt{\langle P_{E}(I) \rangle_{I}\langle P_{S}(I) \rangle_{I}} \right]$. The average deviation between the PDF of a single speckle pattern and the ensemble PDF is $\approx 0.06$ for the family of speckle patterns presented in Figure~\ref{Figure1}. Because this average deviation is the same as what is obtained from an equivalent calculation using Rayleigh speckles, we conclude that our speckle intensity PDFs are statistically stationary. To verify that the intensity PDFs are ergodic, we compare the intensity PDFs of different spatial locations $P_{x} (I)$, with respect to the ensemble PDF $P_{E} (I)$, using $\Delta P_{x} = \left[ \langle |P_{E}(I) - P_{x}(I)| \rangle_{I}\right] / \left[ \sqrt{\langle P_{E}(I) \rangle_{I}\langle P_{x}(I) \rangle_{I}} \right]$. To calculate $P_{x} (I)$ we use the ensemble of intensity values at a given position $x$. For the family of speckles in Figure~\ref{Figure1}, the average deviation between the PDF of a single spatial location and the ensemble PDF is $\approx 0.06$. Again, since this is equal to the value obtained from an equivalent ensemble of Rayleigh speckles, we can conclude that our intensity PDFs are ergodic: in addition to being stationary. 

Similarly for the encoded non-local correlations, one can perform an analogous calculation comparing the spatial intensity correlation function obtained from averaging over all positions in each customized speckle pattern to that obtained from sampling over the ensemble of speckle patterns: to verify that the encoded correlations are stationary. Additionally, one can compare the average spatial intensity correlation function of each spatial position to that obtained from sampling all positions: to verify that the encoded correlations are ergodic. We have checked both cases and for each the average deviation was the same as the value obtained from an equivalent ensemble of Rayleigh speckle patterns. Thus, the intensity correlations encoded into the speckle patterns are both stationary and ergodic. In general therefore, our customized speckle patterns are represented by a statistically stationary and ergodic random process.

\begin{figure}[hthb]
    \centering
    \includegraphics[width= \linewidth]{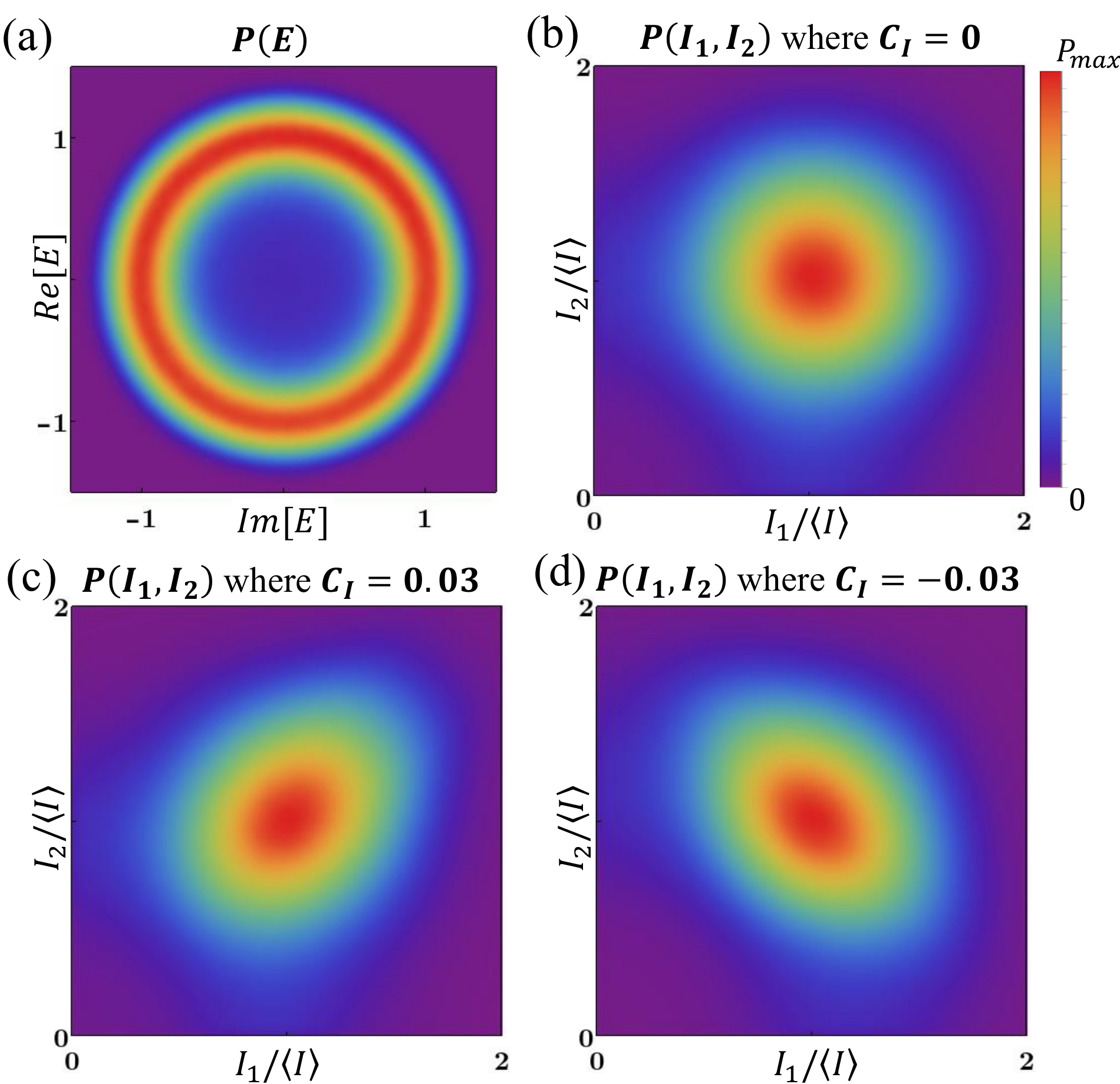}
    \caption{The high-order statistical properties of the family of customized speckle patterns in Fig.~\ref{Figure1} are presented. In (a) the complex joint PDF of the speckle field reveals that the speckle patterns are circular non-Gaussian and therefore fully developed. In (b-d) we show the joint intensity PDF, $P(I_{1}, I_{2})$, of $I_{1}$ and $I_{2}$ sampled at spatial locations separated by $\Delta {\bf R}  = (60$ \textmu m, 60 \textmu m$)$, $(0, 100$ \textmu m$)$, and $(100$ \textmu m, 0$)$ respectively. In (b) the intensity values are uncorrelated and thus the joint intensity PDF is independent, while in (c,d) the encoded non-local correlations result in a dependent joint PDF. To obtain these results we use an ensemble of 5000 customized speckle patterns.}
    \label{Figure2}
\end{figure}

In Figure \ref{Figure2}, we present some of the high-order statistical properties of the family of customized speckle patterns presented in Figure~\ref{Figure1}. We show the complex joint PDF of the speckle field, $P(Re[E], Im[E])$, in Figure~\ref{Figure2} (a). Because the complex field PDF is circular non-Gaussian, we know that the customized speckle patterns are fully developed. Therefore, this indicates that (i) the phases of the speckle fields are uniformly distributed from $0$ to $2 \pi$; (ii) the amplitude and phase values in the complex field are uncorrelated. In (b-d) we show the joint intensity PDF, $P(I_{1}, I_{2})$, of two intensity values, $I_{1}$ and $I_{2}$, separated by a distance , $\Delta {\bf R}$. In (b) we choose the spatial separation, $\Delta {\bf R}  = (60$ \textmu m, 60 \textmu m$)$, at which the spatial intensity correlation function is zero. Because the intensity values are uncorrelated, the joint intensity PDF is independent, $P(I_{1},I_{2})= P(I_{1}) P(I_{2})$. Conversely, when the spatial separation of $I_{1}$ and $I_{2}$ is chosen such that the intensity values are either positively correlated, $\Delta {\bf R}  = (0, 100$ \textmu m$)$, or negatively correlated, $\Delta {\bf R}  = (100$ \textmu m, 0$)$, the joint intensity PDF is dependent, $P(I_{1},I_{2}) \neq P(I_{1}) P(I_{2})$, as shown in Figure \ref{Figure2} (c) and (d). 

While the method presented in this section, to customize the speckle patterns, provides a prescription for creating complex-light fields and controlling statistical properties, it does not provide theoretical limitations on what intensity PDFs and spatial correlations can be realized, which will be addressed in the next section.

\section*{Theoretical Limitations}
 
 \begin{figure*}[hthb]
    \centering
    \includegraphics[width= 14cm]{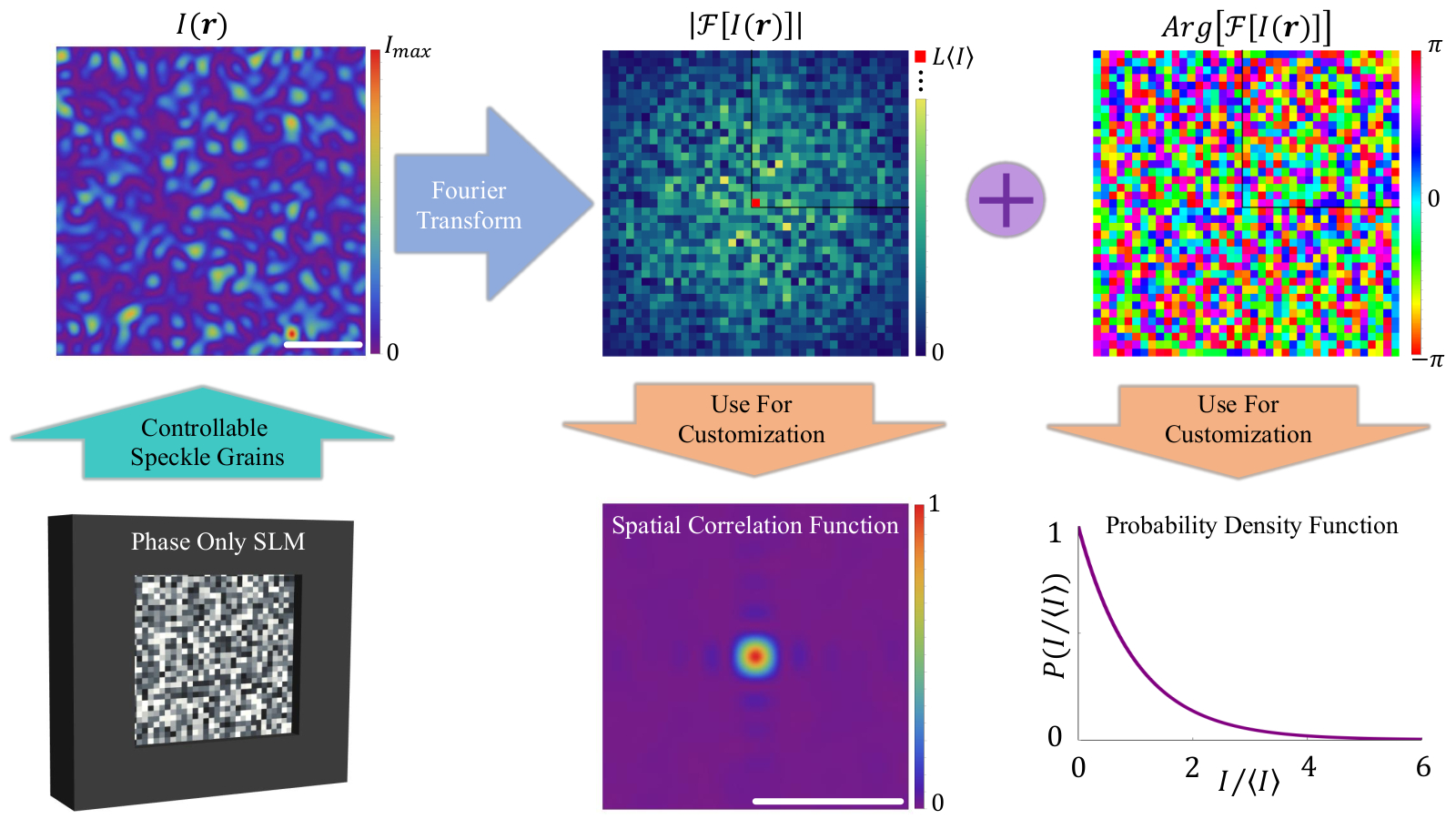}
    \caption{A flowchart illustration of separate control over the speckle correlations and intensity PDF is shown.  The phase-only SLM generates the speckle pattern $I(r)$, whose Fourier transform is $J(\rho) =  \mathcal{F}[I(\mathbf{r})]$. The Fourier amplitude, $|J(\rho)|$, is modulated to manipulate $C_{I}(\Delta\mathbf{r})$ and the phase, ${\rm Arg}[J(\rho)]$, can be used to tailor the intensity PDF.}
    \label{Figure3}
\end{figure*}

In this section, we discuss the degree to which $ C_{I} (\Delta {\bf r})$ and $P(I)$ can be controlled simultaneously and independently in a speckle pattern. To study this, we switch to the Fourier basis of the speckle {\it intensity} $I({\bf r})$; which is distinct from the Fourier transform relationship between the {\it fields} on the SLM and CCD camera planes. Therefore our analysis is general and independent of the precise mapping between the fields on the SLM and camera planes.

For simplicity, we restrict our theoretical calculation to one dimension and consider the speckle intensity pattern consisting of $N$ speckle grains, which can be described by a discrete array with length $L$. We require that $I(r)$ is statistically stationary and ergodic; therefore, spatial averaging over $I(r)$ and ensemble averaging over different $I(r)$ are equivalent processes. Additionally, we normalize our speckle patterns such that $\langle I(r) \rangle$ = 1. Finally, under our conventions the discrete Fourier transform of the speckle pattern can be defined according to:
\begin{equation}
\begin{split}
\label{FT}
\mathcal{F}[I(r)] = J(\rho)  =   \frac{1}{\sqrt{L}} \sum^{L-1}_{r=0} I(r) {\rm e}^{-{\rm i} \frac{2 \pi r \rho}{L}},\\
\mathcal{F}^{-1}[J(\rho)] = I(r)  =   \frac{1}{\sqrt{L}} \sum^{L-1}_{\rho=0} J(\rho) {\rm e}^{{\rm i} \frac{2 \pi r \rho}{L}}.
\end{split}
\end{equation}
In our method of customizing speckle intensity statistics, the spatial correlations are encoded into the speckle pattern, $I(r)$, via modification of the Fourier amplitude of the intensity pattern, $|J(\rho)|$. According to the discrete Wiener-Khinchin theorem, this relationship can equivalently be written as
\begin{equation}
\label{CI_JI}
\begin{split}
\langle I(r)I(r+\Delta r) \rangle =\frac{1}{L} \sum^{L-1}_{\rho=0} |J(\rho)|^{2} \cos\left(\frac{2 \pi \Delta r \rho}{L}\right),\\
|J(\rho)|^{2}=\sum\limits^{L-1}_{\Delta r=0} \langle I(r) I(r+\Delta r)\rangle \cos\left(\frac{2 \pi\Delta r\rho}{L}\right).
\end{split}
\end{equation}
The above equations demonstrate that in a speckle intensity pattern, there is a uniquely determined relationship between the Fourier amplitudes, $|J(\rho)|$ and the intensity correlations $\langle I(r)I(r+\Delta r) \rangle$. Specifically, we know that $\langle I(r)I(r+\Delta r) \rangle$ is what sets $|J(\rho)|$. Therefore for a given desired/arbitrary correlation function, $\langle I(r)I(r+\Delta r) \rangle$, there is a corresponding Fourier amplitude profile, $|J(\rho)|$, which is well defined and can be used to encode the desired/arbitrary correlations into a speckle pattern.

While we cannot directly write the intensity PDF, $P(I)$, as a function of $I(r)$; we can use the relationship between $P(I)$ and its intensity moments, $ \langle I^{n}(r)\rangle  = \int P(I) I^n d I$ where $ n$ is a positive integer. We relate $\langle I^{n} (r) \rangle$ to the Fourier amplitudes of $I(r)$, using the Fourier transformation relation in Eq.~\ref{FT}:
\begin{equation}
\langle I^{n}(r)\rangle = {\bigg \langle} \prod_{k=1}^{n} {\bigg [} \sum_{\rho_{k}=0}^{L-1} \frac{J(\rho_{k})}{\sqrt{L}} {\rm e}^{{\rm i} \frac{2 \pi r \rho_{k}}{L}} {\bigg ]} {\bigg \rangle_{r}}.
\end{equation}
This expression can be simplified by using the delta function identity $\langle  {\rm e}^{{\rm i} \frac{2 \pi}{L} (r_{1}-r_{2})} \rangle_{r_1} = \delta(r_{1},r_{2})$, and written as
\begin{equation}
\label{I_N}
\langle I^{n}(r)\rangle = \frac{1}{L^{n/2}} \sum_{\rho_{1}...\rho_{n-1}=0}^{L-1} J(\rho_{1})... J(\rho_{n-1}) J^{*}(\rho_{1}+...+\rho_{n-1})
\end{equation}
which is valid for $n \geq 2$. For the first moment $n=1$, it can be shown $\langle I(r) \rangle=J(0) / \sqrt{L}$. Eq.~\ref{I_N} shows that the $n^{\rm th}$ moment of a speckle pattern is related to the $n-1$ order correlations among the elements of $J(\rho)$. Comparison between Eqs.~\ref{I_N} and \ref{CI_JI} reveals that while $C_{I}(\Delta r)$, $\langle I \rangle$, and $\langle I^{2} \rangle$ are determined by the amplitude profile of $J(\rho)$, the higher-order intensity moments $n > 2$ can be manipulated separately using the phase values of $J(\rho)$. Since the only relationship between the intensity moment and spatial correlation function is $ \langle I^{2} \rangle =C_{I}(0) + 1$, the $\Delta r$ dependence of the spatial intensity correlation function and the intensity PDF of a speckle pattern can be controlled independently in a speckle pattern: as we illustrate in Figure~\ref{Figure3}. 

While we have established a relative independence between $P(I)$ and $C_{I}(\Delta r)$, this does not necessitate that both can be arbitrarily customized. Next, we identify the limitations on our ability to manipulate $P(I)$ in a speckle pattern. The ability to arbitrarily control the intensity profile of a speckle pattern with $N$ speckle grains is equivalent to controlling the moments of the speckles’ intensity PDF according to Eq.~\ref {I_N} wherein the highest moment one can control is on the order of the number of speckle grains $N$.

The next question is `how many moments are required to uniquely define a PDF?' To answer this question in the context of realistic speckle intensity patterns, it is useful to take certain experimental facts into consideration. For example, all speckle patterns have a finite valued total power (spatially integrated intensity), which imposes a limit on the maximum intensity that a speckle pattern can have, $I_{M}$. Hence, the intensity PDF is bounded by the maximal intensity value $I_{M}$. Furthermore, a measured speckle pattern inherently has discrete intensity values, with the discretization step determined by either the dynamic range of the camera or the measurement noise, and therefore the intensity PDF of the speckle pattern must also have discrete intensity values. Consequently, we define $\Delta I $ as the intensity discretization step, $\Delta I =I_{m+1}- I_{m}$, and as a result the intensity PDF is given by a set of values $P(I_m)$, where $m = 1, 2, ... , M$. The integral equation relating the intensity moments to the PDF can therefore be written in discrete form as $\langle I^{n} \rangle = \Delta I \sum^{M}_{m=1} I_{m}^{n} P(I_{m})$. This relationship is expressed as the following matrix operation:
\begin{equation}
\label{Matrix}
\begin{pmatrix}
\langle I \rangle \\
\langle I^{2} \rangle \\
\vdots\\
\langle I^{N} \rangle
\end{pmatrix}
=\Delta I
\begin{pmatrix}
    I_{1} & I_{2}  & \dots  & I_{M} \\
    I_{1}^{2} & I_{2}^{2}  & \dots  & I_{M}^{2} \\
    \vdots & \vdots  & \ddots & \vdots \\
    I_{1}^{N} & I_{2}^{N}  & \dots  & I_{M}^{N}
\end{pmatrix}
\begin{pmatrix}
P(I_{1}) \\
P(I_{2}) \\
\vdots\\
P(I_{M})
\end{pmatrix}.
\end{equation}
From this relation, reconstructing $P(I)$ from a given number of moments becomes a matrix inversion problem. For the case where $N=M$, the matrix inverse in Eq.~\ref{Matrix} is well defined and can be directly calculated. Therefore, any continuous intensity PDF which can be perfectly represented by $M$ discrete data points is uniquely defined by its first $M$ intensity moments. We can further this line of reasoning beyond the case where $\Delta I$ is dictated by the effective dynamic range of the measurement, and consider the case where $\Delta I$ is the minimum sampling rate required to accurately represent an intensity PDF. The Nyquist-Shannon sampling theorem -which establishes the minimum number of evenly spaced data points required to represent a continuous bandlimited-function without loss of information- tells us that we need to sample $P(I)$ with at least two data points per period of its highest frequency component: in order to represent the continuous bandlimited-function $P(I)$ with a discrete array, $P(I_{m})$, without losing any information. This means that if $P(I)$ can be represented by a band limited and Nyquist-sampled array of $M$ data points, then $P(I)$ is uniquely defined by its first $M$ intensity moments. In other words, if in a speckle pattern we can uniquely control $M$ intensity moments, then the speckle pattern can possess any intensity PDF which is band limited and satisfies the Nyquist sampling theorem when discretized into $M$ data points.

\begin{figure}[hthb]
    \centering
    \includegraphics[width= \linewidth]{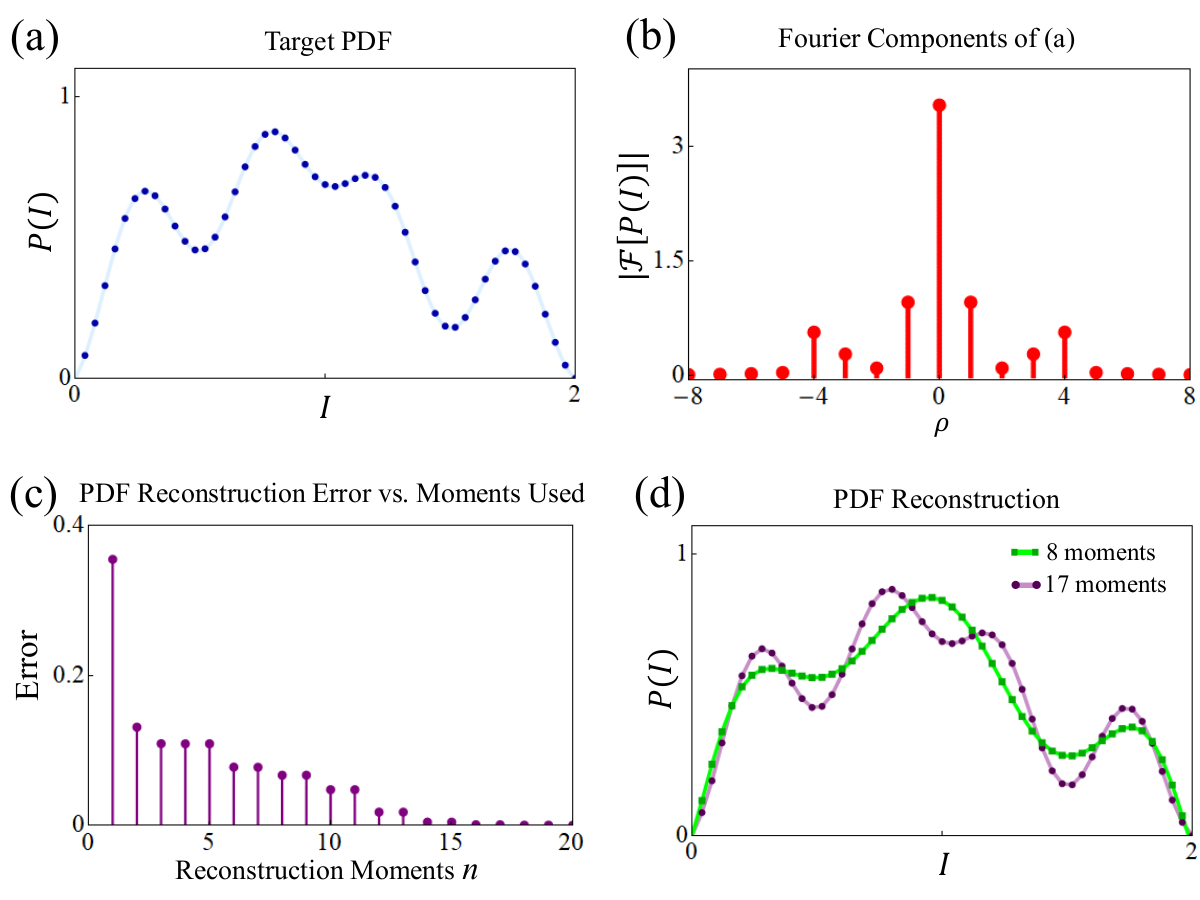}
    \caption{We show the effect of reconstructing an intensity PDF from differing numbers of its moments. An example PDF, $P(I)$, with a finite range of intensity is shown in (a). The Fourier spectrum of $P(I)$ is plotted in (b), demonstrating that the PDF has a limited number of non-zero Fourier components. (c) shows the PDF reconstruction error of (a) as a function of the number of intensity moments used for reconstruction. Two examples of reconstructed PDFs are shown in (d); they are created using 8 intensity moments and 17 intensity moments.}
    \label{TheoryFig3}
\end{figure}

In Figure~\ref{TheoryFig3} we demonstrate this principle using the example PDF shown in (a). In this example we set $I_{M}=2$ and $\Delta I =1/25$,  thus $M = 50$, as seen in (a). In (b) the amplitude of the bandlimited Fourier spectrum of $P(I)$ is shown, after $P(I)$ has been Nyquist sampled. In this case the Fourier amplitudes are non-zero between $-4 \leq \rho \leq 4$ and zero for $4 < |\rho |\leq 8$. Hence, at least 17 intensity moments are required to uniquely define the $P(I)$ shown in (a): in term of its intensity moments. We can verify this by applying the pseudo-inverse of the $M \times N$ matrix in Eq. ~\ref{Matrix} to different numbers of moments of the PDF: under the condition that the number of moments used is less than the number of data points $N < M$. In this case, when we have a sufficient number of moments to uniquely define the PDF in (a), at least 17 moments, the PDF reconstructed from its moments is identical to the original. In (c) we plot the PDF reconstruction error of (a) as a function of the number of intensity moments $N$ used in the reconstruction. Around when the number of moments reaches the critical number to uniquely define the PDF, the error vanishes. In (d), we plot two example PDFs reconstructed from $N= 8$ and $N=17$ moments. While the reconstruction using $N=8$ intensity moments fails to reproduce the correct PDF, increasing $N$ to $N=17$ results in a faithful reproduction of the original PDF.

In the context of using a phase-only SLM in 1D, if we modulate $2N$ independent phase-values on the SLM, we can control the intensity profile of $N$ speckle grains, and therefore the total degree of control we have on the complex $J(\rho)$ values is $N$. Half of it, $N/2$, lies in the amplitude of $J(\rho)$ which is used to manipulate spatial correlations $C_{I} (\Delta {\bf r})$. The other half, lies in the phase of $J(\rho)$ and translates to the ability to control $N/2$ intensity moments. As long as a desired intensity PDF  can be Nyquist-sampled and represented in bandlimited form by less than $N/2$ data points, a speckle pattern adhering to it can be generated. In 2D, the only practical difference is that we require $4N$ independent phase values to control a speckle pattern with $N$ speckle grains. At this point, it is important to recall that we require speckle patterns to be stationary and ergodic. In order to satisfy these conditions, the speckle patterns must consist of a large number of speckle grains, because the average difference between the ensemble PDF of a family of speckle patterns and a single realization scales as $1/\sqrt{N}$. Therefore, $N$ is always large in practice, \textit{e.g.}, $N\approx 250$ in our experiments. For realistic intensity PDFs, however, the number of non-zero Fourier components is much less that $N/2$ due to noise limitations,\cite{CSS} and therefore they can be reconstructed with significantly less than the available $N/2$ intensity moments. To summarize, we have found that in a speckle intensity pattern the intensity PDF and the functional form of $C_{I}(\Delta r)$ can be controlled independently and arbitrarily except for the constraint $C_{I}(0)= \langle I^{2} \rangle  -1$.

\begin{figure}[hthb]
    \centering
    \includegraphics[width= \linewidth]{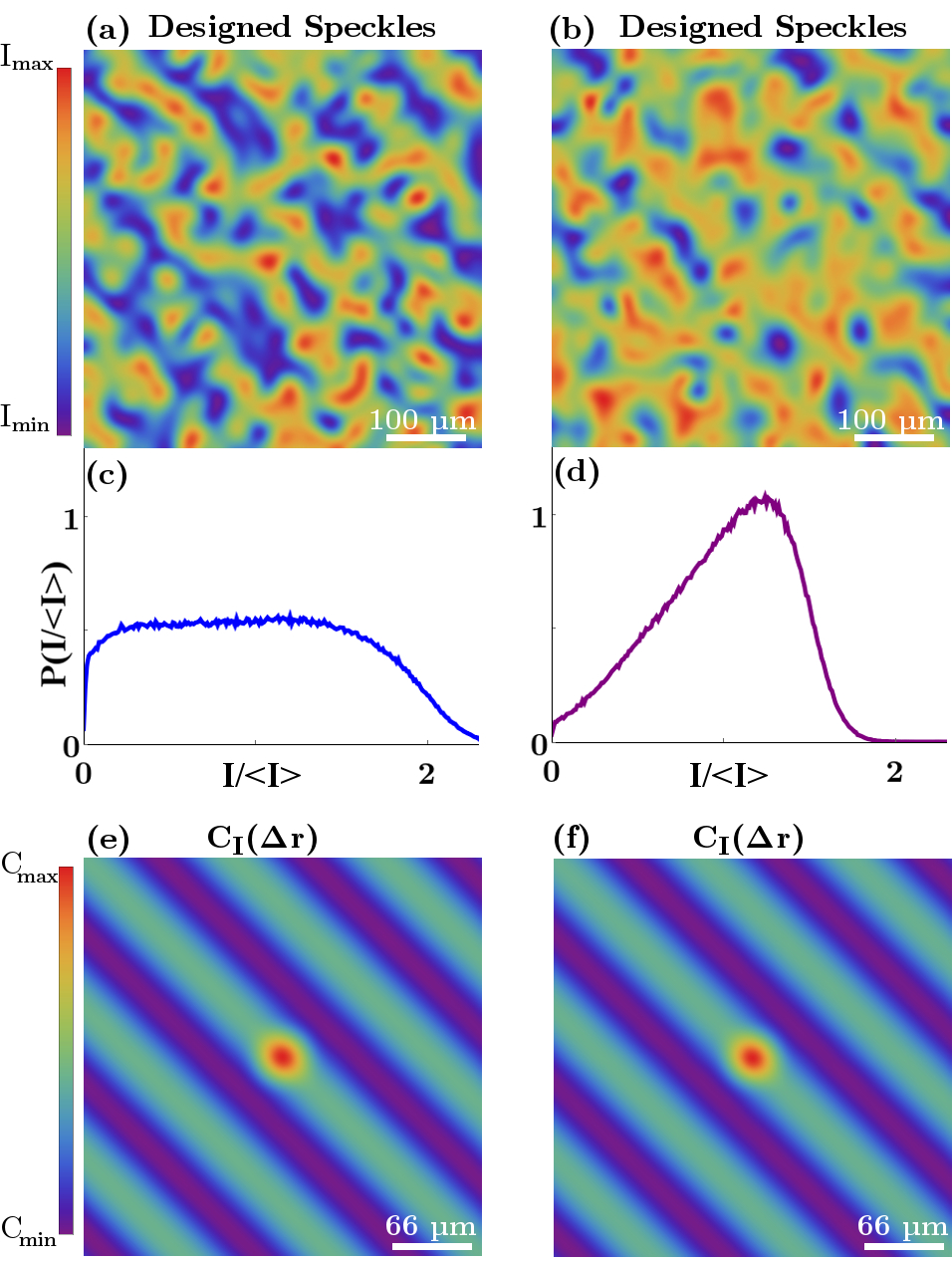}
    \caption{Two customized speckle patterns (a,b) with different intensity PDFs (c,d) but identically shaped spatial intensity correlation functions are shown in (e,f). In (a) we present a measured speckle pattern adhering to both the PDF shown in (c), which is flat over a pre-defined range of $I/\langle I \rangle$, and the diagonally oscillating spatial intensity correlation function shown in (e). In (b) we present a measured speckle pattern with the same shaped intensity correlation function (f), but obeying a different intensity PDF (d), which increases linearly over a pre-defined $I/\langle I \rangle$ range. In (e) $C_{\rm max}=0.32$ and $C_{\rm min}=-0.09$ while in (f) $C_{\rm max}=0.14$ and $C_{\rm min}=-0.04$ due to the different PDFs. We ensemble average over 100 independent speckle patterns to obtain the PDF and correlation functions in (c-f). The origin of (e) and (f) is located at the center of each plot.}
    \label{Figure6}
\end{figure}

\section*{Experimental Realizations}

\begin{figure}[hthb]
    \centering
    \includegraphics[width= \linewidth]{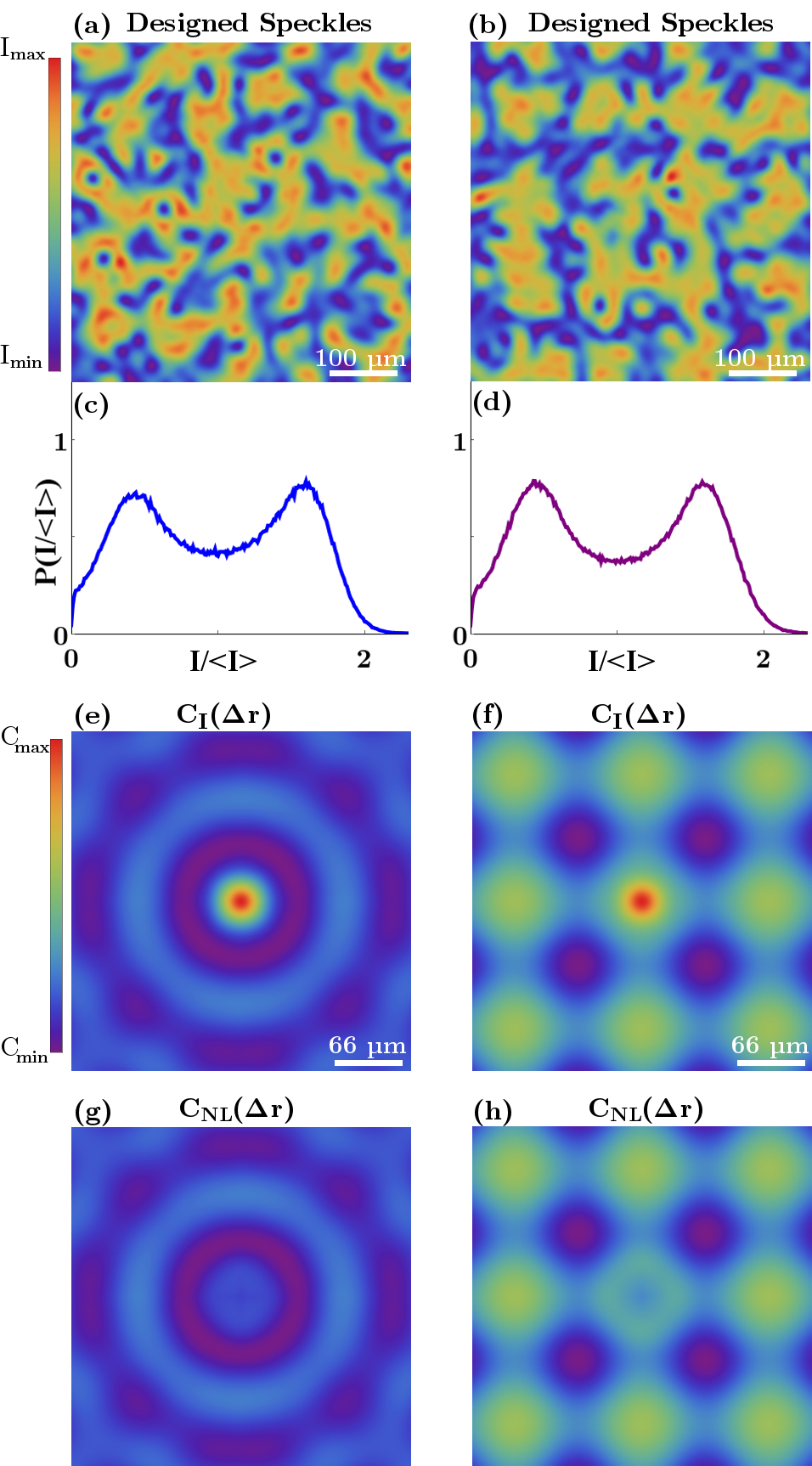}
    \caption{Two customized speckle patterns with the same intensity PDF but different spatial intensity correlation functions are shown. In (a) we present a measured speckle pattern adhering to both the bimodal intensity PDF shown in (c) and the isotropic oscillating correlation function shown in (e). In (b) we present a measured speckle pattern also adhering to a bimodal PDF (d) but possessing anisotropic `checkerboard' correlations shown in (f). In (g,h) we show the corresponding non-local correlations. In (e) $C_{\rm max}=0.29$ and $C_{\rm min}=-0.05$ while in (f) $C_{\rm max}=0.30$ and $C_{\rm min}=-0.12$. We ensemble average over 100 independent speckle patterns, to obtain the PDFs and correlation functions in (c-h). The origins of (e-h) are located at the center of each plot.}
    \label{Figure7}
\end{figure}

Next we experimentally demonstrate an independent control over the intensity PDF and the spatial intensity correlations of a speckle pattern. In Figure~\ref{Figure6} we show two examples of speckle patterns that obey different $P(I)$, but have congruent $C_{I}(\Delta {\bf r})$. The intensity correlation function is designed to have non-local correlations of the form , $C_{NL}(\Delta {\bf r}) = (2 C_{I}(0)/7) \cos[(\Delta {\bf x} + \Delta {\bf y})/{20}]$, where $x$ and $y$ are spatial coordinates. For the example customized speckle pattern in (a), the intensity PDF is designed to be constant, $P(\hat{I})=1/2$, over the intensity range $0 \leq \hat{I} = I/ \langle I \rangle \leq 2$ and zero elsewhere. The second example speckle pattern, shown in (b), is designed to obey a different intensity PDF which linearly increases, $P(\hat{I})=\hat{I}$, over the intensity range $0 \leq \hat{I} \leq \sqrt{2}$. As a result of obeying different intensity PDFs, $\langle \hat{I}^{2} \rangle$ differs between the two speckle patterns and therefore $C_{I}(0) = 0.32$ in (e) while $C_{I}(0) = 0.14$ in (f). While there is a visible difference in the speckle contrast between (a) and (b) due to the different intensity PDFs, the spatial intensity correlations in both speckle patterns have the same functional form: as can be seen in (e) and (f). Comparison between (a) and (b) illustrates how the topology of the customized speckle patterns changes in accordance with the PDF, while the overarching spatial order is dictated by the non-local correlations. Both speckle patterns in (a,b) have an overarching diagonal oscillation. Nevertheless, for the speckles with a linearly increasing PDF in (b), the spatial intensity profile has the appearance of an interwoven web of bright channels with randomly dispersed dark islands, while the speckles with a uniform PDF lack any definite channel structure beyond the diagonal oscillations.

In Figure~\ref{Figure7} we show two examples of speckle patterns that have the same intensity PDF, however, their spatial intensity correlation functions differ. The speckle patterns in both (a) and (b) adhere to a bimodal intensity PDF, as shown in (c) and (d). The local intensity transformation which generates speckles with the PDF, $P(\hat{I})=\sin^{2}(\pi \hat{I})$ over the intensity range $0 \leq \hat{I} = I/ \langle I \rangle \leq 2$ and zero elsewhere, is used to create these speckle patterns. However, because optical fields must be continuous functions, $P(\hat{I}) \neq 0$ over $0 < \hat{I} < I_{\rm M}$, and therefore the experimental PDFs deviate from $P(\hat{I})=\sin^{2}(\pi \hat{I})$ around $\hat{I} =1$ as originally shown in. \cite{CSS} Despite having the same intensity PDF, the non-local intensity correlation functions of (a) and (b) are designed to have different spatial variations. In (e) the spatial intensity correlation function $C_{I}(\Delta {\bf r})$ is an azimuthally-symmetric radially-oscillating function with the appearance of a ‘bullseye’. The non-local correlation function, shown in (g), is designed to have the form: $C_{NL}(\Delta {\bf r}) = (C_{I}(0)/6) \sin[(\Delta r )/{14}]$. In contrast, in (f) $C_{I}(\Delta {\bf r})$ is designed to be an anisotropic function having a `checkerboard' form with non-local correlations, (h), of the form: $C_{NL}(\Delta {\bf r}) = [4 C_{I}(0)/10] \cos[(\Delta {\bf x} + \Delta {\bf y})/{40}]\cos[(\Delta {\bf x} - \Delta {\bf y})/{40}]$. While both speckle patterns in (a, b) share a similar topology, consisting of two interlaced bright and dim channels, the overarching structure differs in both its shape and orientation. In (b) the checkerboard correlations induce the formation of multi-speckle islands with a grid-like orientation. In (a), the bullseye correlations result in an interwoven web-like structure.

\section*{Conclusion and Discussion}

In conclusion, we have experimentally demonstrated a method of customizing the intensity probability density functions of speckle patterns; while simultaneously introducing long-range spatial correlations among the speckle grains. The customized speckle patterns exhibit radically different topologies and varying degrees of spatial order. In addition to our experimental demonstration, we have explored both the theoretical and practical limitations on the extent to which the intensity probability density function and the spatial intensity correlations can be manipulated separately and arbitrarily in a speckle pattern. 

Although the camera is placed on the Fourier plane of the SLM in our experiment this is not a necessity. Our method can easily be adapted to customize the statistical properties of speckle patterns on other $2D$ planes: or even when a random scattering medium is placed in between the SLM and camera. To accomplish this, we simply need to measure the field-transmission matrix which maps the field on the SLM surface to the field incident on our camera. Therefore, our method and theoretical description provide a systematic approach for creating complex light fields and controlling their statistical properties with a phase-only spatial light modulator: while also providing the upper bounds on what is possible.

There are numerous avenues of research, related to customizing speckle patterns, that are worth further exploration. For example, our method tailors the speckle patterns at a specific plane defined by the camera. Such patterns, similar to Rayleigh speckle patterns, exhibit a rapid axial-decorrelation: occurring within the Rayleigh range of an optical system. In addition to decorrelating within one Rayleigh range, our customized speckles lose both their tailored intensity PDFs as well as their non-local correlations as they axially propagate away from the target plane. \cite{CSS, IntroNonLocal} Whether or not it is possible to control the statistical properties of speckle patterns simultaneously on multiple planes -or even in a 3D volume- remains an open question. \cite{leonetti2015observation}

Finally, we will discuss some of the potential applications of our customized speckle patterns. Because our method of creating and controlling complex light is versatile -yet simple- it can readily be incorporated into an extensive range of optical applications and experiments. For example, the ability to arbitrarily control the non-local correlations and intensity PDFs of speckle patterns could enhance many structured-illumination applications like speckle illumination microscopy, \cite{Dynamic2, Dynamic4, 2019_Guillon_NatCommun} super-resolution imaging, \cite{Super_opt1,  Super_acous1} and high-order ghost imaging. \cite{image1,image2, image3} Similarly, it could also benefit studies of cold atom,\cite{cold} active media, \cite{active} and microparticle \cite{coll} transport in correlated optical potentials. Our method is advantageous because both the topology and the degree of spatial order in the speckled optical potentials are arbitrarily customizable and reconfigurable without any mechanical motion.

\section*{Acknowledgments}
This work is supported by the MURI grant no. N00014-13-1-0649 from the US Office of Naval Research.

\bibliography{Bib}

\end{document}